\documentclass[aps,prd,preprint,floatfix,nofootinbib]{revtex4}
\usepackage{graphicx}

\begin{document}



\title{Novel Collider Signatures for Little Higgs Dark Matter Models}

\renewcommand{\thefootnote}{\fnsymbol{footnote}}

\author{ 
Chian-Shu Chen
\footnote{Email address: d927305@oz.nthu.edu.tw}, 
Kingman Cheung
\footnote{Email address: cheung@phys.nthu.edu.tw}, 
and Tzu-Chiang Yuan
\footnote{Email address: tcyuan@phys.nthu.edu.tw} }
\affiliation{Department of Physics and NCTS, National Tsing Hua University, 
Hsinchu, Taiwan}

\renewcommand{\thefootnote}{\arabic{footnote}}
\date{\today}

\begin{abstract}
Little Higgs models with $T$-parity provide a stable neutral particle
$A_H$, the lightest $T$-odd particle (LTP), which was recently
proposed to explain the cold dark matter of the Universe.  In the
coannihilation region of the LTP with the Little Higgs partner of the
light quark $Q_H$, we propose novel signatures for its detection at
hadronic colliders by searching for monojet and dijet plus missing 
energy events
due to the associated production of $Q_H A_H$ and direct production
of $Q_H \overline{Q}_H$, respectively.
Using the most recent WMAP data, we show that the coupling of 
$A_H$-$Q_H$-$q$ is tightly constrained in the coannihilation region. 
This implies that the monojet signal is too small to be discernible but
the dijet plus missing energy signal is potentially measurable at the
LHC.
\end{abstract}

\maketitle
 
\section{Introduction}
Recent studies on cosmology have almost unanimously confirmed the
existence of dark matter (DM) in our Universe with a fraction
$\Omega_{DM} \simeq 0.22$ of the critical density \cite{wmap}.
However, the identity of the DM still remains to be a mystery.  Future
DM direct and indirect detection experiments and collider experiments
will attempt to pin down the properties of the DM candidates and the
underlying theories.  Among various proposed candidates in the
literature, the most studied one is the lightest supersymmetric
particle (LSP, usually the lightest neutralino) 
of supersymmetric theories (SUSY).  Recently, another class
of models based on Little Higgs models \cite{summary} imposed with a
$T$-parity \cite{T1,T2,T3} provides an alternative DM candidate.  In
these models, all the standard model particles are assigned with an
even $T$-parity, while all the new particles proposed are assigned
with an odd $T$-parity, except for the Little Higgs partner of the top
quark. Because of the $T$-parity assignment the lightest $T$-odd
particle (LTP) is stable and hence a potential candidate to account
for the DM of the Universe \cite{asano,martin,lhdm}.  We note that
electroweak precision measurements have placed severe constraints on
earlier attempts of constructing the Little Higgs models without
$T$-parity \cite{csaki}.  
Imposing a discrete $T$-parity in the Little Higgs models
therefore opens a much wider window for theorists to explore various
phenomenologies of this class of models \cite{pheno}.

In Little Higgs models with $T$-parity,
the LTP is usually the $A_H$, the corresponding heavy Little Higgs partner 
of the $B$ field of the $U(1)_Y$ symmetry.
\footnote{The $A_H$ and $Z_H$ fields are the mass eigenstates of $B_H$
and $W_H^0$ fields after a rotation by an angle analogous to the
Weinberg angle, but the rotation in the present case is highly
suppressed by $v/f$, where $f \sim 1$ TeV is the high scale where
$[SU(2) \times U(1)]^2 \rightarrow SU_L(2) \times U_Y(1)$ in the
Little Higgs models and $v \sim 246$ GeV is the VEV to break $SU_L(2)
\times U_Y(1) \rightarrow U_{em}(1)$.  Thus, $A_H$ is very close to
$B_H$.  }
It has been shown \cite{asano,lhdm}
that in order for the LTP to account for all the DM of the Universe, the
mass $M_{A_H}$ of $A_H$  is related to the Higgs boson mass $m_H$ by 
\cite{lhdm}
\begin{equation}
  m_H \approx 24 + 2.38 M_{A_H} \qquad {\rm or} \qquad 
  m_H \approx -83 + 1.89 M_{A_H} \;.
\end{equation}
This mass relation between $m_H$ and $M_{A_H}$ clearly indicates that
the dominant annihilation channel is via an $s$-channel Higgs boson:
$A_H A_H \to h^{(*)} \to b\bar b,\; \tau^+ \tau^-, \; W^+ W^-, ZZ$.
Twice the mass of $A_H$ must be located at either side of the Higgs
boson peak, $2 M_{A_H} \approx m_H + \Delta$.  If $2 M_{A_H}$ is located right
at the Higgs boson mass, the annihilation would be too strong to
provide a sufficient relic density for the DM.  Therefore, $2 M_{A_H}$
must falls on either side of the Higgs boson mass such that the
annihilation cross section has just the right size to produce the
correct relic density for DM.  This is somewhat a fine-tuned mechanism.
Perhaps a slightly more natural domain of the parameter space is a
region where the coannihilation of the LTP ${A_H}$ and the $T$-odd
partner of light quark $Q_H$ becomes strong.  This
coannihilation region has been explored and is given by \cite{lhdm}
\begin{equation}
M_{A_H} + 20 \approx M_{Q_H} \;\; ,
\end{equation}
with $M_{Q_H} \geq 200$ GeV.  Note that no mass relation is necessary
for $m_H$ and $M_{A_H}$ in this coannihilation region.  
In this work, we consider that the
coannihilation region is a more natural setting and we focus on the
physics of this interesting region.

\begin{figure}[t]
\centering
\includegraphics[width=5in]{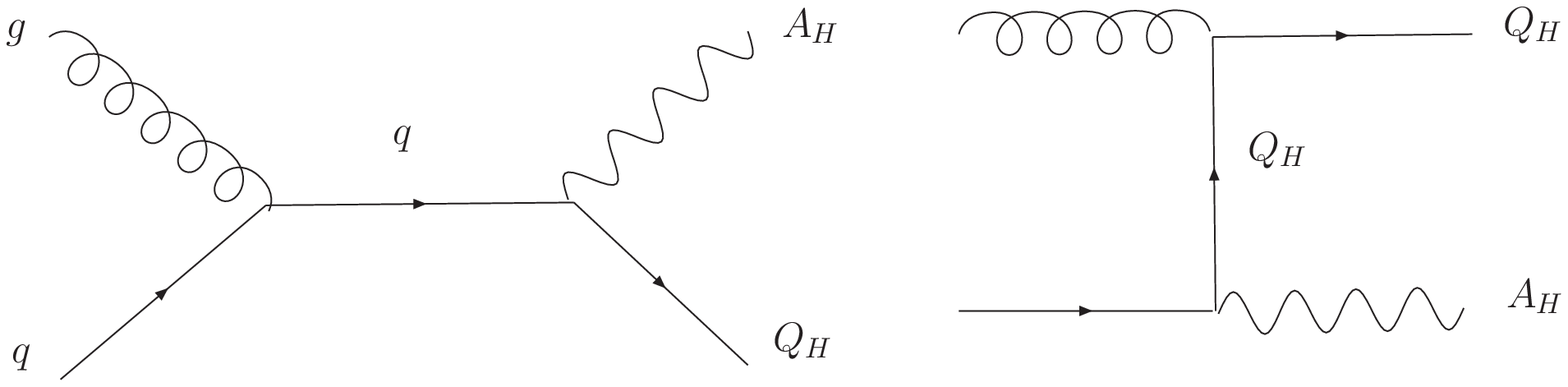}
\includegraphics[width=5in]{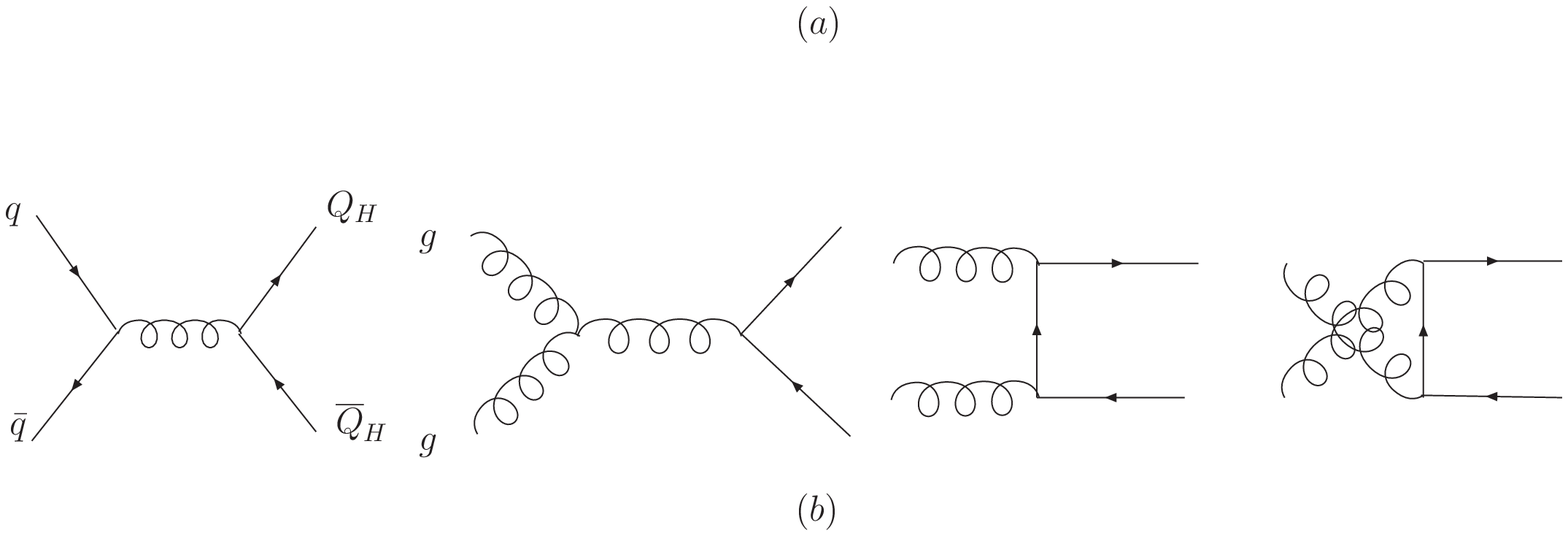}
\caption{\label{fig1}
Feynman diagrams for (a) production and coannihilation of $A_H Q_H$ and
(b) production and coannihilation of $Q_H \overline{Q}_H$.
Production reads from left to right, whereas coannihilation reads 
from right to left.}
\end{figure}

The Feynman diagrams responsible for coannihilation is depicted in 
Fig. \ref{fig1}. 
It has been widely discussed in the literature that the LHC is a good place
to investigate the nature of the DM candidate.  In this work,
we propose novel signatures for the LTP of the Little Higgs models with
$T$-parity in the coannihilation
region.  We consider the associated production of $pp \to A_H Q_H$,
followed by the decay $Q_H \to q A_H$, which will give rise to 
monojet plus missing energy events, as well as the direct production
of $Q_H \overline{Q}_H$, which gives rise to 
the dijet plus missing energy signal.

One can argue that using monojet plus missing energy events
one can at least partially distinguish Little Higgs models with
$T$-parity from minimal SUSY models with $R$-parity.  Some of the
reasons are (i) monojet events arising from 
$gq \rightarrow {\tilde q}\widetilde{\chi}^0_1$, 
followed by ${\tilde q} \rightarrow \widetilde{\chi}^0_1 q$ with 
the lightest neutralino $\widetilde{\chi}^0_1$ as the LSP, 
is expected to be suppressed by the 
coupling. Thus, the dominant channels for producing SUSY particles are
QCD production of $\tilde{g}\tilde{g}$, $\tilde{q} \tilde{q}^*$, and
$\tilde{g}\tilde{q}(\tilde{q}^*)$.  They tend to give multijet ($> 2$)
plus missing energy events;  (ii) the dominant mechanism for producing
$T$-odd particles are $q g \to A_H Q_H$, $\bar q g \to A_H \overline{Q}_H$,
  and 
$gg,q\bar q \to Q_H \overline{Q}_H$, 
which give rise to monojet and dijet, respectively, plus
missing energy events.  Therefore, by counting how many jets in
multijet plus missing energy events, one may be able to differentiate
between Little Higgs models with $T$-parity and minimal SUSY models.
Note that it is possible to have some corners of parameter space 
in minimal SUSY
models such that monojet plus missing energy events become 
discernible in SUSY, 
but the chance is very odd albeit not zero.
\footnote{If the mass difference between the squarks and 
the lightest neutralino is large, the monojet and dijet arising from SUSY 
tend to have a harder $p_T$ spectrum than those coming from the $Q_H$ 
decays in the coannihilation region of the Little Higgs models.}

Our analysis of relic density in Sec. III shows that the coupling of
$A_H$-$Q_H$-$q$ has to be rather small, such that the monojet plus missing
energy signal is negligible compared with the Standard Model (SM)
background.  Nevertheless, the dijet plus missing energy signal via
direct $Q_H \overline{Q}_H$ production is comparable with the SM
background.  Potentially, the dijet signal is feasible at the LHC.

We organize our paper as follows.  In the next Section, we list the
formulas for the parton processes involved in our calculation.  
In Sec. III, we use the most recent WMAP result of the constrained DM 
cross section to obtain a suitable range for the coupling of $A_H$-$Q_H$-$q$. 
In Sec. IV, we calculate the monojet and dijet plus missing energy signals 
at hadronic colliders.  Comparison of the predictions with the SM 
backgrounds are also discussed in this Section.
We conclude in Sec. V.

\section{Parton processes for production of  
$A_H Q_H$ and $Q_H \overline{Q}_H$}
The relevant interaction Lagrangian density for the vertex 
$A_H$-$Q_H$-$q$ is given by 
\cite{lhdm,han-logan-mcelrath-wang}
\begin{equation}
{\cal L} = g_H \, \overline{Q}_H \, \gamma_\mu \, P_L \,q \, A_H^{\mu}  
\;\; + \;\; {\rm H.c.}\;,
\end{equation}
where the color indices for $Q_H$ and $q$ are suppressed, and $g_H$ is
the coupling constant to be determined. The QCD interaction of $Q_H $
with the gluon fields is the same as that of the light quark $q$.

\subsection{$q g \to A_H Q_H$ and $\bar q g \to  A_H \overline{Q}_H$}

There are two Feynman diagrams, depicted in Fig. \ref{fig1}(a), contributing
to this process.  Summing over the spins and colors of the initial and 
final state particles, the amplitude
squared for the process $q(p_1)\; g(p_2) \rightarrow A_H(k_1) \;Q_H(k_2)$
is given by
\begin{eqnarray}
{\sum} | M|^2 &=& 8 g_H^2 g_s^2 \Biggr\{
  -\frac{1}{\hat s} \left(  \hat t_Q + \frac{\hat s_{AQ} \hat u_A}{M_{A_H}^2}
  \right ) 
   \nonumber \\
&& -\frac{1}{\hat t_Q^2} \biggr[
 2 \hat t_A (\hat s_{AQ} + \hat u_Q) +12 M_{Q_H}^2 \hat t_A - 
  (M_{A_H}^2 - M_{Q_H}^2) \hat u_Q
 \nonumber \\
&&  + \frac{\hat t_A}{M_{A_H}^2} ( 4 M_{Q_H}^2 \hat t_A + \hat u_Q \hat t_A 
+ M^2_{Q_H} \hat s_{AQ} ) 
   \biggr ] \nonumber \\
&&  -
\frac{1}{\hat s \hat t_Q} \Biggr[  \left( 1 + \frac{\hat t_A}{M_{A_H}^2} 
\right )
   ( \hat u_Q \hat u_A - \hat t_A \hat t_Q + \hat s \hat s_{AQ} ) 
  - 4 \hat u_Q(\hat s+\hat t_A + \hat u_A) \nonumber \\
  && - 2  M_{Q_H}^2 \left( 2\hat s - \frac{\hat t_A}{M_{A_H}^2}
   (\hat t_A +\hat u_A ) \right ) \Biggr ]  \Biggr \}   \;\;\; ,
 \label{amp}
\end{eqnarray}
where $\hat s = (p_1 + p_2)^2$, $\hat t = (p_1 - k_1)^2$, 
$\hat u = (p_1 - k_2)^2$, and 
$\hat s + \hat t + \hat u = M_{Q_H}^2 + M_{A_H}^2$. 
We have also introduced the following notations:
$\hat t_Q = \hat t - M_{Q_H}^2$, $\hat t_A = \hat t - M_{A_H}^2$, 
$\hat u_Q = \hat u - M_{Q_H}^2$, $\hat u_A = \hat u - M_{A_H}^2$, 
and $\hat s_{AQ} = \hat s - M_{A_H}^2 - M_{Q_H}^2 $.
In Eq.(\ref{amp}), $g_s$ is the strong coupling constant and 
$g_H = g' {\tilde Y}$ with $g'$ and $\tilde Y$ defined in \cite{lhdm}.
The differential cross section for this subprocess is then given by
\begin{equation}
\frac{d \hat \sigma}{d \cos\theta^*} = 
\frac{\beta}{3072 \pi} \, \frac{1}{\hat s} \, \sum |M|^2 \;,
\end{equation}
where $\theta^*$ is the scattering angle in the parton center-of-mass frame
and $\beta = [ ( 1 - M_{Q_H}^2/\hat s - M_{A_H}^2/\hat s)^2 - 4 
 ( M_{Q_H}^2/\hat s )(M_{A_H}^2/\hat s) ]^{1/2}$. 
Similar expression can be written down for 
 $\bar q g \to A_H \overline{Q}_H$ as well.

\subsection{$g g , q \bar q \to Q_H \overline{Q}_H$}

The QCD production of the $Q_H \overline{Q}_H$ pair is similar to the 
case of the top-quark pair.  Feynman diagrams are depicted in 
Fig. \ref{fig1}(b).
Hence the cross sections can be easily adapted from previous calculations.
The  differential cross sections for these subprocesses 
are given by \cite{km}
\begin{eqnarray}
\frac{d\hat\sigma}{d \hat t} (q\bar q \to Q_H \overline{Q}_H) 
  &=& \frac{8\pi \alpha_s^2}{9 \hat s^2} \, 
\left[ \frac{1}{2} - v +z \right]  \; ,
 \\
\frac{d\hat\sigma}{d \hat t} (gg \to Q_H \overline{Q}_H) 
  &=& \frac{\pi \alpha_s^2}{12 \hat s^2} \, \left( \frac{4}{v} - 9 \right )\,
   \left [ \frac{1}{2} - v + 2z \left( 1- \frac{z}{v} \right )\right ]
 \;,
\label{ggQQ}
\end{eqnarray}
where $z = M_{Q_H}^2/\hat s$ and $v= (\hat t -M_{Q_H}^2 ) (\hat u -M_{Q_H}^2 )
 / \hat s^2$.
In the above cross section for $q\bar q \to Q_H \overline{Q}_H$, we
have neglected the contribution from an $A_H$ exchange in the
$t$-channel. Compared with the QCD diagram, this contribution is
suppressed by the weak coupling 
$g_H^2$  
as well as the heavy mass
effect in the propagator of $A_H$.  The above cross sections of
various parton-level subprocesses are convoluted with the parton
distribution functions to obtain the monojet and dijet cross sections
at hadron colliders.  We will return to this at Section IV.

\section
{Coannihilation cross sections and the Coupling $g_H$}

The most recent WMAP data \cite{wmap} can give us some crude
estimation of the coupling $g_H$. We will focus at the coannihilation
region where $M_{Q_H}$ is close to the LTP mass $M_{A_H}$. The
relevant annihilation processes are $A_H A_H \rightarrow f \bar f$
with $f$ stands for either a lepton or a quark, $Q_H \overline{Q}_H
\rightarrow q \bar q, gg$, and $A_H Q_H (\overline{Q}_H) \rightarrow q
(\bar q) g$. Their matrix elements can be easily obtained and are
listed as below.

\subsection{$A_H(k_1) A_H(k_2) \rightarrow f(p_1)  {\bar f}(p_2):$}

\begin{eqnarray}
\overline{\sum} \vert  M \vert^2 & = & {1 \over 9} g^2_H N^2_c 
\Biggl\{ {1 \over {\hat t}_Q^2} \left[ \hat t(7\hat t+4 \hat u)  - 4 M_{A_H}^4 
- 4 {\hat t^2(\hat t + \hat u) \over M^2_{A_H}} + 
{\hat t^3 \hat u \over M_{A_H}^4} \right]  \nonumber \\
&& + {1 \over {\hat u}_Q^2} \left[ \hat u(7\hat u+4\hat t) -4 M_{A_H}^4 
- 4 {\hat u^2(\hat t+\hat u) \over M_{A_H}^2} + 
{\hat t \hat u^3 \over M_{A_H}^4}  \right] \nonumber \\
&&  - {2 \over {\hat t}_Q {\hat u}_Q} 
\left[ 7 \hat t \hat u +8 M_{A_H}^2 (\hat t + \hat u) - 16 M_{A_H}^4
- 4{\hat t \hat u( \hat t + \hat u) \over M_{A_H}^2} + 
{\hat t^2 \hat u^2 \over M_{A_H}^4} \right] \Biggr\}
\;\; ,
\label{AAff}
\end{eqnarray}
where $N_c = 1$ or $3$ for $f$ equals a lepton or a quark, respectively. 
Here the subscript label $Q$ can be referred 
to the heavy lepton or quark depends on whether $f$ is a lepton or a quark.

\subsection{$Q_H(k_1) \overline{Q}_H(k_2) \rightarrow q(p_1)  {\bar q}(p_2):$}
\begin{eqnarray}
\overline{\sum} \vert M \vert^2 & = &
\left( {1 \over 2 N_c} \right)^2 \Biggr\{ 8 g_s^4  {(N_c^2 - 1)^2 \over 4} 
{1 \over \hat s^2}
\left[ \hat t_Q^2 + \hat u_Q^2 + 2 M_{Q_H}^2 \hat s \right] \nonumber \\ 
&& +g_H^4  N_c^2 {1 \over \hat t_A^2} \left[ 4 (\hat s + \hat t_Q)^2 
+ 4 {M^4_{Q_H}\hat s  \over M^2_{A_H}} + 
{M^4_{Q_H}\hat t^2_Q  \over M^4_{A_H}} \right] \nonumber\\
&& - 2 g_H^2 g_s^2 {N^2_c - 1 \over 2}  {1 \over \hat s \hat t_A} 
\left[ 4 (\hat s + \hat t_Q)^2 + 4 M^2_{Q_H}\hat s  + 
2 {M^2_{Q_H} \over M^2_{A_H}} 
\left( \hat t^2_Q + M^2_{Q_H} (\hat s + \hat t) \right) \right] \Biggr\}  \; .
\label{QQqq}
\end{eqnarray}

\subsection{$Q_H(k_1)  \overline{Q}_H(k_2) \rightarrow g(p_1)   g(p_2):$}
\begin{eqnarray}
\overline{\sum} \vert M \vert^2 & = &
\left( {1 \over 2 N_c} \right)^2 {64 \over 3} g_s^4 \left( 1 -  4 z \right)
 \left( {4 \over v} - 9 \right) 
\left[ {1 \over 2} - v + 2 z \left( 1 - {z \over v} \right) \right] \;\; ,
\label{QQgg}
\end{eqnarray}
where $z$ and $v$ are defined as those after Eq.(\ref{ggQQ}).

\subsection
{$A_H(k_1)  Q_H(k_2) \rightarrow q(p_1)  g(p_2)$ and 
$A_H(k_1)  \overline{Q}_H(k_2) \rightarrow \bar q(p_1)  g(p_2)$ 
}
\begin{eqnarray}
\overline{\sum} \vert M \vert^2 & = & {1 \over 3} {1 \over 2 N_c} 
{\rm Eq.}(\ref{amp}) \;\; .
\label{AQqg}
\end{eqnarray}

\subsection{Estimation of $g_H$}

A useful formula for estimating the relic density of a 
weakly-interacting massive particle is given by
\cite{hooper}
\begin{equation}
  \Omega_{\chi} h^2 \approx \frac{ 0.1\;{\rm pb}}{\langle \sigma v \rangle}\;.
\label{usefulformula}
\end{equation}
The most recent result from WMAP is \cite{wmap}
\begin{equation}
\Omega_{\rm CDM} h^2 = 0.105 \pm 0.009 \;,
\label{wmapdata}
\end{equation}
where we have used the WMAP-data-only fit and taken $\Omega_{\rm CDM} =
\Omega_{\rm matter} - \Omega_{\rm baryon}$.  
Using Eq.(\ref{usefulformula}), one can translate this WMAP data
to $\langle \sigma v \rangle$
\begin{equation}
\langle \sigma v \rangle \approx 0.95 \pm 0.08 \; {\rm pb} \;.
\end{equation}
The effective annihilation cross section due to coannihilation can be
written as
\[
 \langle \sigma_{\rm eff} v \rangle = \sum_{ij} \langle \sigma_{ij} v_{ij}
 \rangle \, \frac{n^{\rm eq}_i n^{\rm eq}_j} { n^2} \; ,
\]
where $n = \sum_i n^{\rm eq}_i$ and the equilibrium number density 
$n_i^{\rm eq}$ of each species $i$ is given by
\[
  n^{\rm eq}_i \sim g_i \left (\frac{m_i T}{2\pi}\right)^{3/2} e^{-m_i/T} \;,
\]
where $g_i$ is the spin degree of freedom of the species $i$.
The relative velocity $v_{ij}$ is 
\[
v_{ij} = \frac{\sqrt{(p_i \cdot p_j)^2 - m^2_i m^2_j}}{ E_i E_j} \;,
\]
where $E_i$ and $p_i$ are the energy and 4-momenta 
of the species $i$, respectively. 
In the center-of-mass frame of $(i,j)$, the relative
velocity $v_{ij}$ in the non-relativistic limit can be simplified to
\begin{equation}
v_{ij} =\frac{ 2 \lambda^{1/2}(1, m_i^2/s, m_j^2/s )}
             { 1 + {\cal O}(m^4_i/s^2)}
      \approx 2  \lambda^{1/2}(1, m_i^2/s, m_j^2/s ) \; ,
\end{equation}
where $\lambda(1,m_i^2/s, m_j^2/s) = ( 1- m_i^2/s - m_j^2/s)^2 - 4 (m_i^2/s)
(m_j^2/s)$. 

\begin{figure}[t!]
\centering
\includegraphics[width=5in]{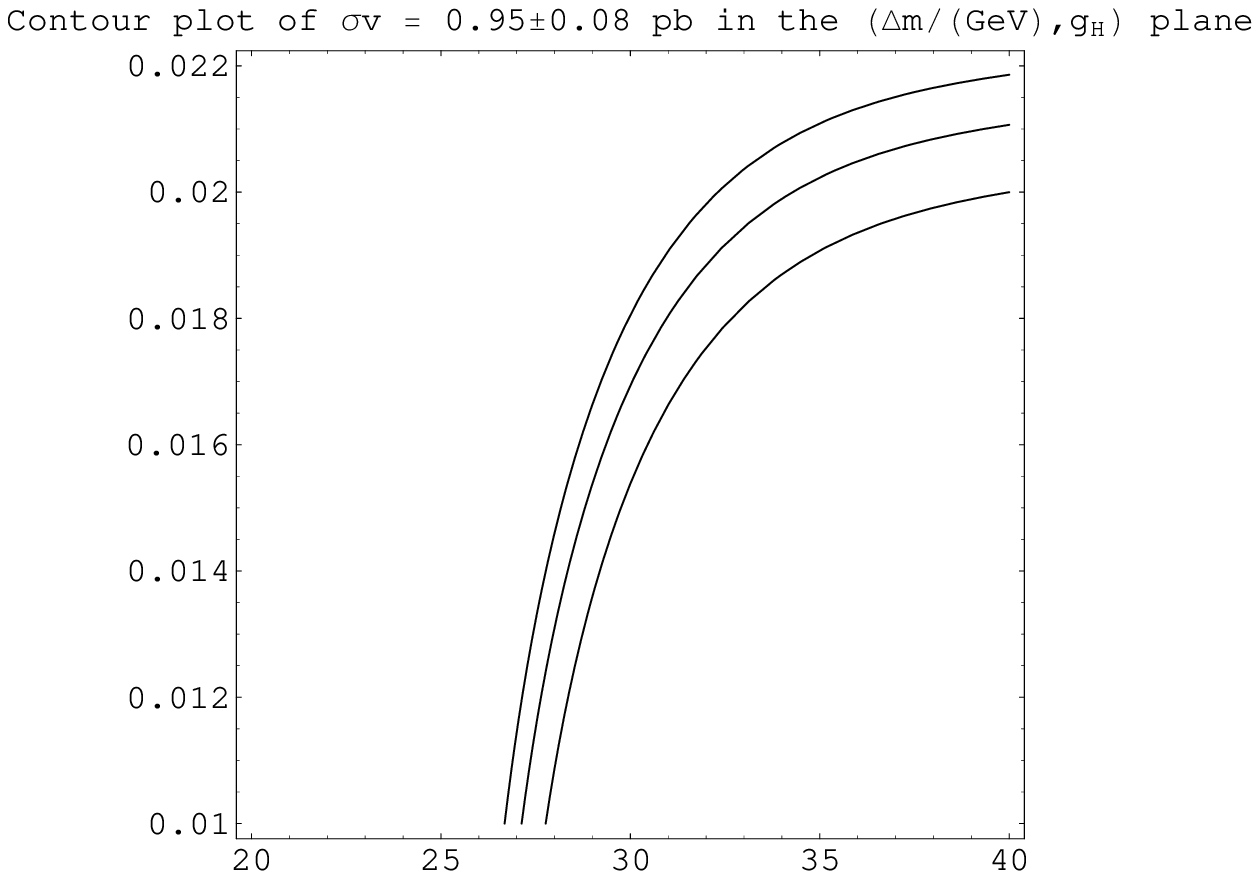}
\caption{\label{fig2}
The contour plot of coannihilation cross section $\sigma_{\rm eff} v$
in the plane of $(\Delta m, g_H)$. We take $M_{A_H}=200$ GeV. 
The WMAP implied result of $0.95 \pm 0.08$ pb for this quantity is shown.}
\end{figure}

In the present coannihilation scenario, we assume that there are 
5 flavors of heavy quarks (from the heavy partners of the 
first two generation quarks and the partner of the $b$-quark of 
the third generation).  
The resultant effective annihilation cross section is given by
 \begin{eqnarray}
 \langle \sigma_{\rm eff} v \rangle &=& \frac{1}{  \left[1 + 
 10\, \frac{g_{Q_H}}{g_{A_H}} \,\left( 1 + \frac{\Delta m}{M_{A_H}} 
   \right)^{3/2} \, e^{-\Delta m/T} \right ]^2 } \; \biggr \{
   5\, \sigma_{A_H A_H \rightarrow q \bar q} \; v_{A_H A_H} \nonumber \\
&& + 10 \,\sigma_{A_H Q_H \rightarrow q g} \; v_{A_H Q_H} \, 
\left( \frac{g_{Q_H}}{g_{A_H}} \right )\,
     \left( 1 + \frac{\Delta m}{M_{A_H}} \right )^{3/2} \, 
     e^{-\Delta m/T} \nonumber \\ 
&& + 5\,\left( \, \sigma_{Q_H \overline{Q}_H \rightarrow q \bar q} 
   + \sigma_{Q_H \overline{Q}_H \rightarrow gg} \right ) 
\;v_{Q_H \overline{Q}_H} \,
  \left( \frac{g_{Q_H}}{g_{A_H}} \right )^2 \,
     \left( 1 + \frac{\Delta m}{M_{A_H}} \right )^{3} \, 
     e^{-2 \Delta m/T}  \biggr \} \;, \label{eff}
\end{eqnarray}
where $\Delta m = M_{Q_H} - M_{A_H}$, 
$g_{Q_H} =2$, $g_{A_H}=3$, and $\sigma_{ij}$
is the annihilation cross section between species $i$ and $j$. 
We have assumed that twice the mass of the LTP $A_H$ is far away enough 
from the Higgs pole, then only the coannihilation processes listed in 
this Section are important.
The coannihilation cross sections can be obtained from the
amplitude-squared listed in Eqs. (\ref{AAff}), (\ref{QQqq}), (\ref{QQgg}),
and (\ref{AQqg}). 
Taking the simple approximations that the freeze-out temperature
occurs at $T_f \approx M_{A_H}/25$ and during the freeze-out the
species have a relative velocity $\beta =\lambda^{1/2}( 1, M_{A_H}^2/s,
M_{Q_H}^2/s) \approx 0.3$, we ignore the thermal average and
evaluate the quantity $\sigma_{\rm eff} v$ directly from
Eqs.(\ref{eff}) and (\ref{AAff})--(\ref{AQqg})
 as a function of three parameters
$g_H$, $\Delta m \equiv M_{Q_H} - M_{A_H}$, and $M_{A_H}$.  In
Fig. \ref{fig2}, we show the contours of $\sigma_{\rm eff} v= 0.95 \pm
0.08$ pb in the plane of $(\Delta m, g_H)$ for the case $M_{A_H}=200$
GeV. From the figure, it is interesting to see that when $\Delta m$
increases along the contour, the required $g_H$ has to 
increase so as to come up with
sufficient coannihilation rate.  For a reasonable $\Delta m \alt 30$ GeV and 
$M_{A_H}$ in the range varies from 200 to 400 GeV,  $g_H$ never gets 
larger than $0.02$.
It is obvious that these values of $g_H$ is too small to achieve a meaningful 
production cross section for $A_H Q_H$ at colliders, 
and thus a negligible signal
of monojet plus missing energy.  We therefore focus on the
dijet plus missing energy signal, which is independent of $g_H$, 
in the next Section.

\section{Hadronic production}

In the coannihilation region, $Q_H$ decays predominantly into $A_H  q$. 
Its decay rate is given by 
\begin{equation}
\Gamma (Q_H \rightarrow A_H  q) = \frac{1}{32\pi} g^2_H M_{Q_H} 
\left( 2 + {M^2_{Q_H} \over M^2_{A_H}} \right) 
\left( 1 - {M^2_{A_H} \over M^2_{Q_H}} \right)^2  \; .
\end{equation}
Since this is the only possible decay mode of $Q_H$ 
in this region, its branching ratio into a jet
plus missing energy is 100\%.  
The decay time calculated using the above
formula is so short that the decay is actually prompt.  
In the rest frame of $Q_H$, the available energy to the jet depends on
the mass difference $M_{Q_H} - M_{A_H} \sim 20$ GeV, which is not
so small. 

In the production of $A_H Q_H$, the final state has only one jet with
missing transverse energy.  In this monojet signal, 
the $p_{T_j}$ of the jet is the same as
the missing transverse momentum $\not\!{p}_T$.  
Although the monojet plus missing energy signal is very clean, there
are still a handful of SM backgrounds one needs to
contemplate.  The major SM background comes from $Z+1j$ production
followed by the invisible decay of the $Z$ boson.  Unfortunately, we cannot
use the recoil mass information of the monojet to reject this $Z+1j$
background in the hadronic environment. The other reducible monojet SM
background comes from, for example, $W+1j \to \ell\nu + 1j$ with the
charged lepton unidentified.  
With the small value of $g_H$ obtained in the last Section, the production
cross section of $A_H Q_H$ is too small compared with the $Z+1j \to 1j +
\not\!{p}_T$ background.  In the following, we focus on the direct production
of $Q_H \overline{Q}_H$ which gives rise to dijet plus missing energy signal.

In the production of $Q_H \overline{Q}_H$, the final state consists 
of two jets with missing transverse energy.  
The corresponding major SM background
is $Z+2j$ production with $Z\to
\nu\bar \nu$.  This SM background has been studied in great details
by experimental collaborations \cite{cdf-d0} and by theoretical
calculations \cite{theory}.  We use MADGRAPH \cite{madgraph}
to generate the helicity amplitudes of the tree-level
$Z+2j$ background, followed by $Z\to \nu \bar \nu$ decay.  
We impose the following acceptance cuts for the jets:
\begin{equation}
p_{T_j} > 20 \; {\rm GeV}\;, \;\;\; |y_j|< 1.5\;, \;\;\;
 {\rm and} \;\;\; \Delta R_{jj} > 0.7 \;, \label{cuts}
\end{equation}
where $p_{Tj}$ is the transverse momentum, $y_j$ is the rapidity, and
$\Delta R = \sqrt{ (\Delta y)^2 + (\Delta \phi)^2 }$ denotes the
spatial separation of the two jets with $\Delta y$ and $\Delta \phi$ 
denote the differences in their rapidities and azimuth angles, respectively.

\begin{figure}[t!]
\centering
\includegraphics[width=5in]{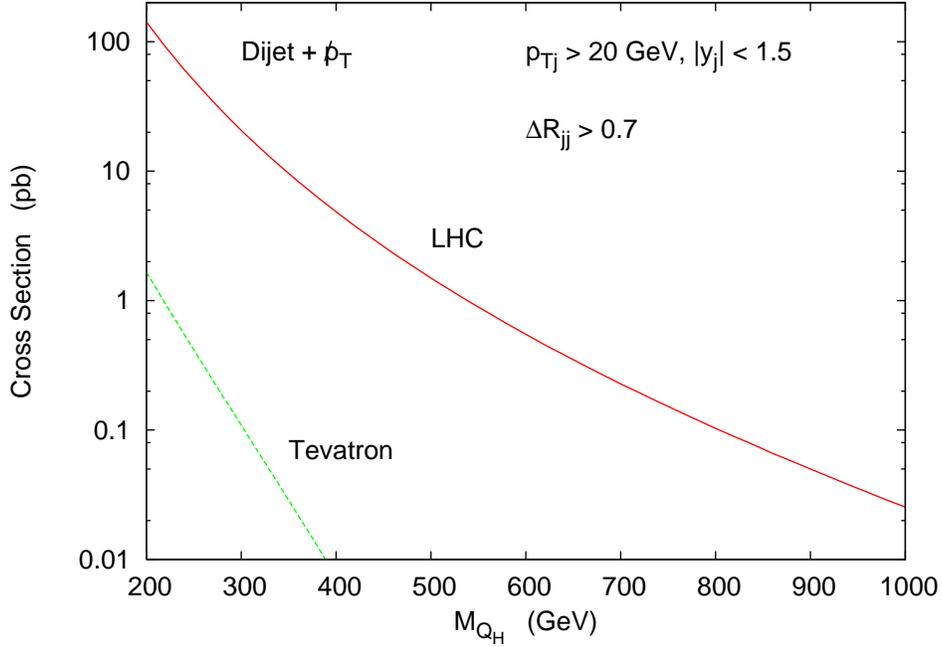}
\caption{\label{fig3} \small
Production rate for the dijet plus missing energy signal
with the acceptance cuts defined in Eq. (\ref{cuts}) at the Tevatron and at
the LHC.}
\end{figure}

\begin{figure}[t!]
\centering
\includegraphics[width=3.2in]{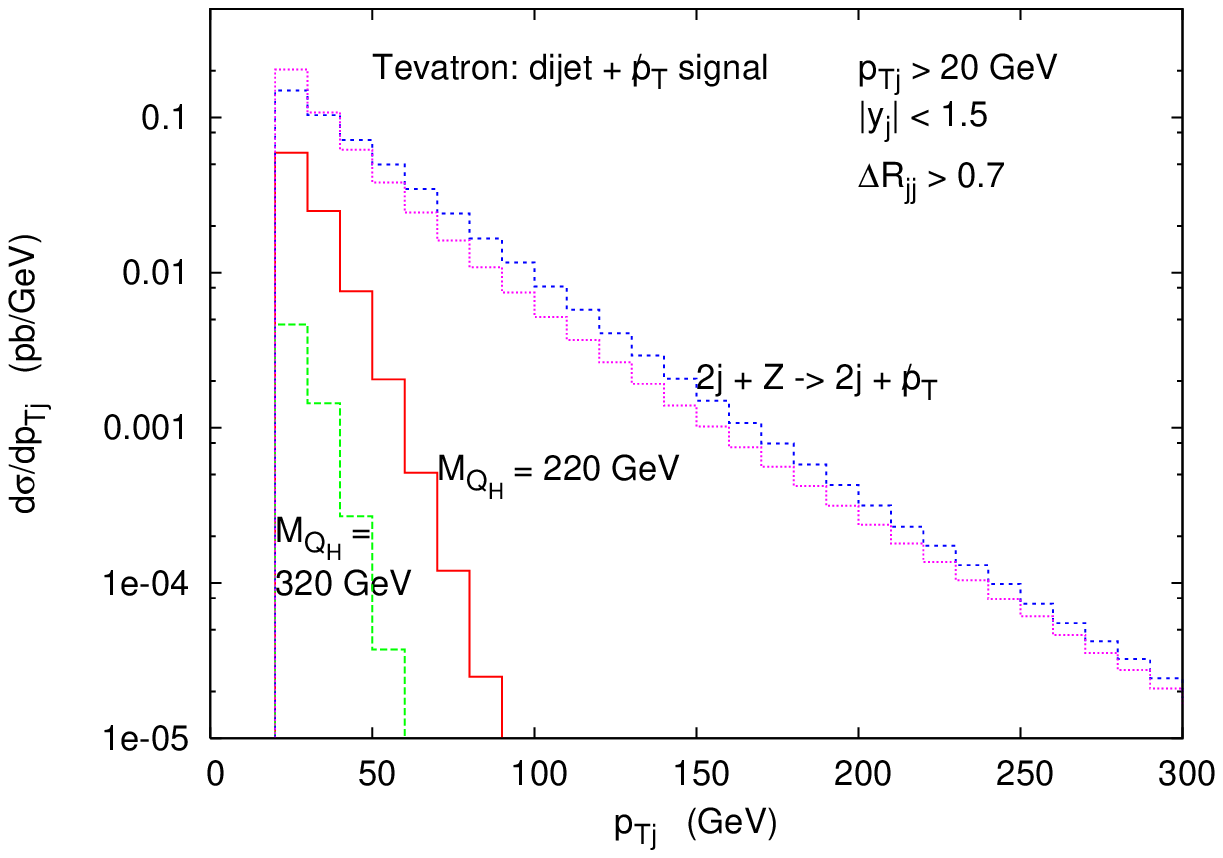}
\includegraphics[width=3.2in]{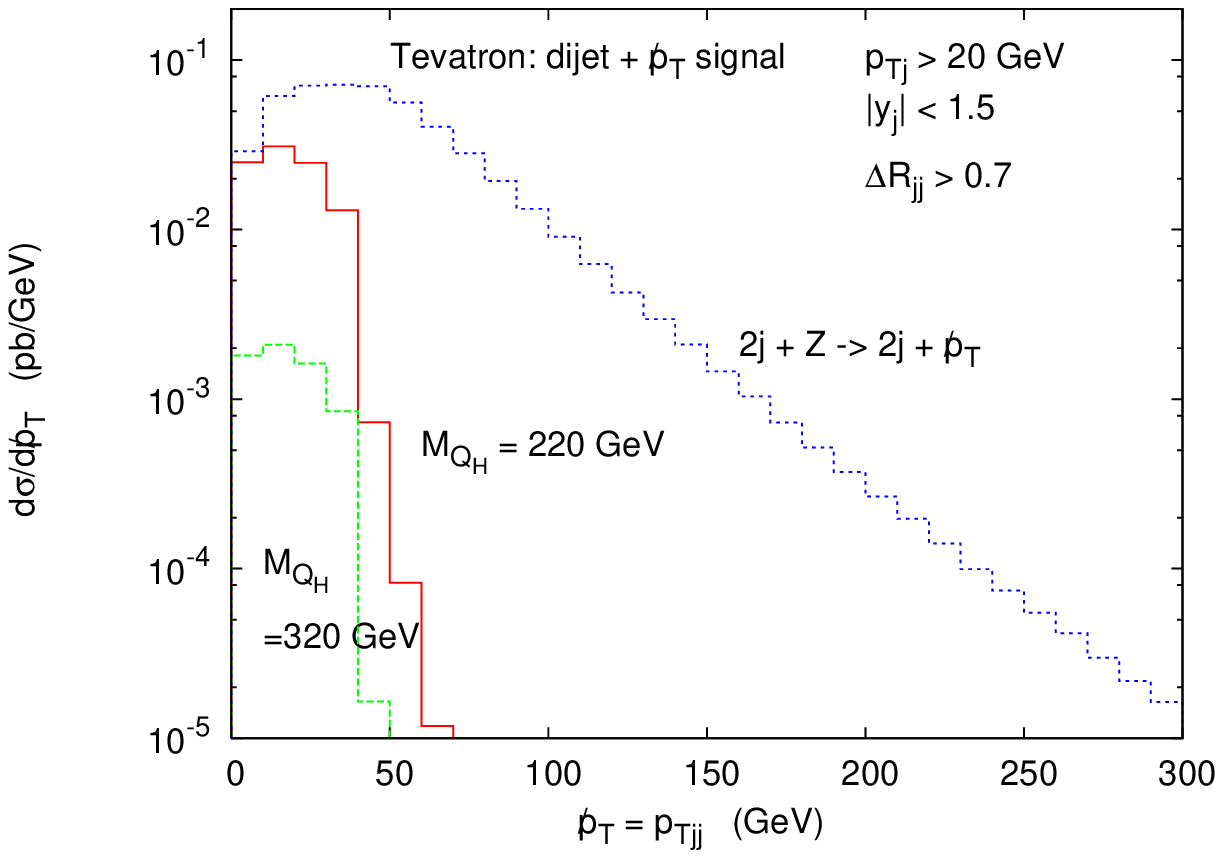}
\caption{\label{fig4} \small
The differential cross section versus (a) the transverse momentum of the jets
and (b) the missing transverse momentum
for the dijet $+ \not\!{p}_T$ signal and the $Z+2j \to 2j + \not\!{p}_T$
background at the Tevatron.  We have used 2 sets of $(M_{Q_H}, M_{A_H})=
(220, 200)$ and $(320,300)$ GeV.
}
\end{figure}

\begin{figure}[t!]
\centering
\includegraphics[width=3.2in]{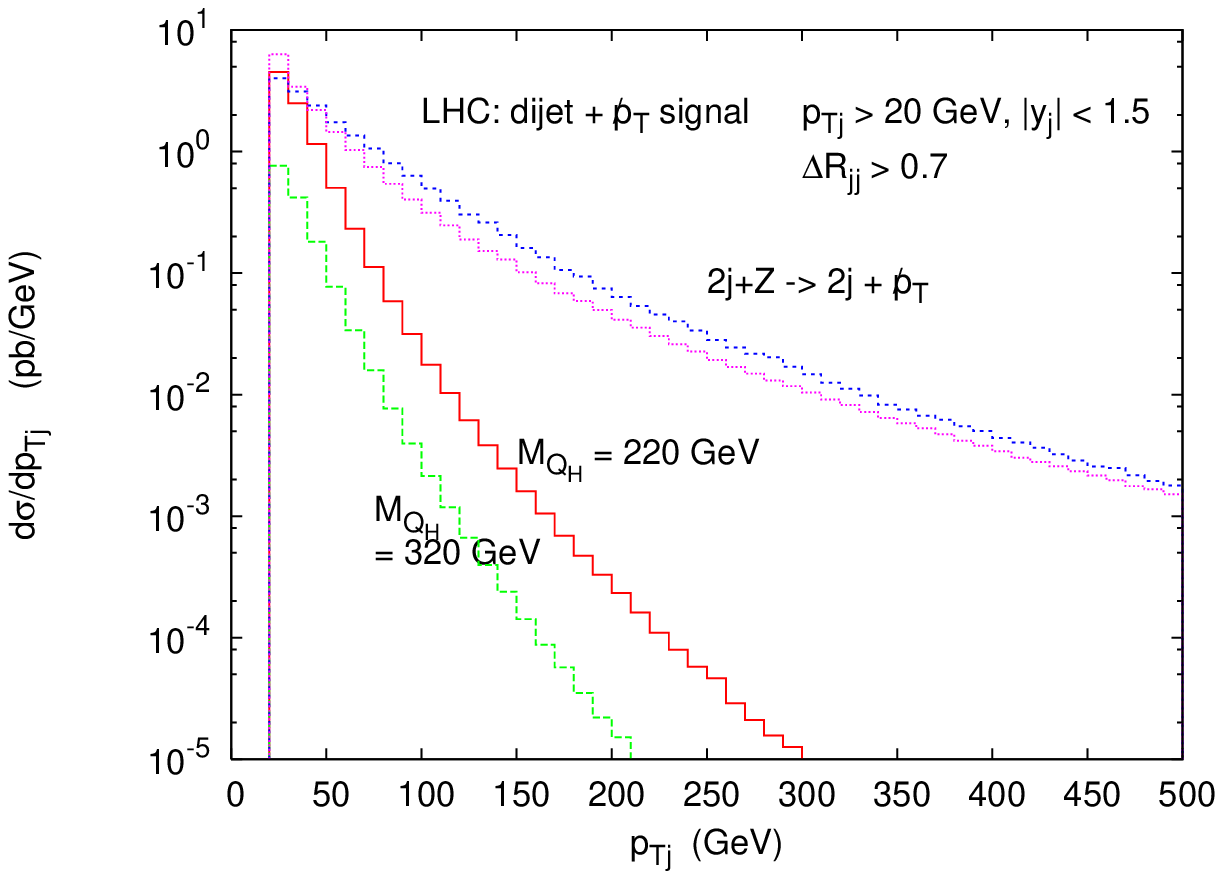}
\includegraphics[width=3.2in]{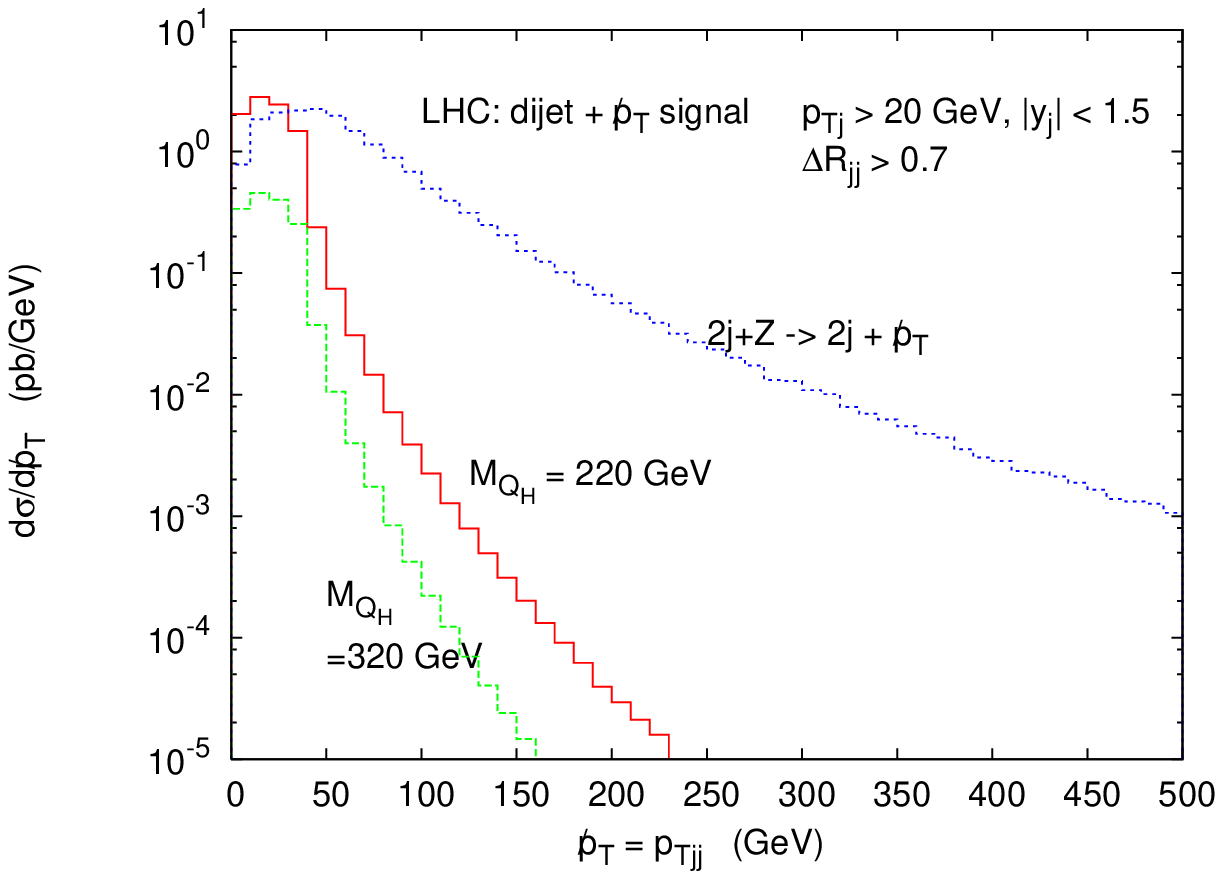}
\caption{\label{fig5} \small
The differential cross section versus (a) the transverse momentum of the jets
and (b) the missing transverse momentum
for the dijet $+ \not\!{p}_T$ signal and the $Z+2j \to 2j + \not\!{p}_T$
background at the LHC.  We have used 2 sets of $(M_{Q_H}, M_{A_H})=
(220, 200)$ and $(320,300)$ GeV.
}
\end{figure}

In Fig. \ref{fig3},
we show the production rate for the dijet plus missing energy signal
with the acceptance cuts defined in Eq. (\ref{cuts}) as a function of 
$M_{Q_H}$.
The spectra of the transverse momentum of the jets and the missing
transverse momentum for the dijet plus missing energy signal and the 
$Z+2j \to 2j + \not\!{p}_T$ background at the Tevatron and at the LHC
are shown in Figs. \ref{fig4} and \ref{fig5}, respectively.
We have included 5 degenerate flavors of $Q_H$ in the signal,
and multiplied the SM background rate by 
the branching ratio $B(Z\to \nu \bar \nu) = 0.2$ to obtain the 
$Z+2j \to 2j + \not\!{p}_T$
background.
It is clear that the spectra for the Little Higgs dark matter model 
are softer than those from  
the background, reflecting the fact that  the mass 
difference $\Delta m = M_{Q_H} -
M_{A_H}$ is relatively small in the coannihilation region. Note that 
this soft region in the missing transverse momentum spectra overlaps 
with the Jacobian peak of the $Z$ from the SM background.
However, the SM $Z+2j$ background can be measured with high accuracy
using $\ell^+ \ell^- + 2j$ mode.  The $Z+2j \to 2j + \not\!{p}_T$ can
then be obtained quite reliably by using both the leptonic and
invisible branching ratios of the $Z$ boson.  One therefore should be
able to quantify the SM background of $2j + \not\!{p}_T$ with high
accuracy.  By measuring the $p_T$ spectra of the jets at the low end
accurately and/or any distortion around the Jacobian peak in the
missing transverse momentum spectra of the dijet precisely, one could
identify the existence of the dijet signal from the Little Higgs dark
matter model at the coannihilation region.

\section{Conclusions}
Little Higgs models with $T$-parity provide the weakly interacting
particle $A_H$ as the lightest $T$-odd particle to be an interesting
alternative candidate for the dark matter in the Universe.  We have
pointed out a potentially novel signature in the 
direct production of $Q_H \overline{Q}_H$ in the Little Higgs models 
with $T$-parity in the
coannihilation region where $M_{Q_H} - M_{A_H} \alt 20$ GeV.  
It gives rise to dijet plus missing energy events at the Tevatron and at
the LHC, which is potentially discoverable despite the SM background
of $Z + 2j \to 2j + \not\!{p}_T$.
Note that the associated production of $Q_H A_H$, which gives rise to 
a cleaner monojet $+\not\! E_T$ signature, is unfortunately very tiny because 
of the small value of the coupling $g_H$ constrained by the WMAP data.

We close by reiterating that one may use the monojet plus missing energy 
signature to at 
least partially distinguish Little Higgs models with $T$-parity from 
the minimal SUSY models.  
SUSY events tend to give multijet
($> 2$) plus missing energy events
from the QCD production of $\tilde{g}\tilde{g}$, $\tilde{q} \tilde{q}^*$, 
and $\tilde{g} \tilde{q}(\tilde{q}^*)$.  
On the other hand, the dominant production for 
$T$-odd particles are $gg,q\bar q \to Q_H \overline{Q}_H$ 
and $gq \to Q_H A_H$, which give rise to dijet and monojet, respectively,
plus missing energy events.  Therefore, by counting how many jets
in multijet plus missing energy events, one may be able to differentiate
between Little Higgs models with $T$-parity and minimal SUSY models.  

\section*{Acknowledgment}
This research was supported in part by the National Science Council of Taiwan
R.~O.~C.\ under Grant Nos.\ NSC 94-2112-M-007-010-,
and by the National Center for Theoretical Sciences.


\end{document}